\documentclass[a4paper,11pt]{article}
\usepackage{geometry}                
\usepackage{graphicx}
\usepackage{amssymb}
\usepackage{epstopdf}
\usepackage{array}
\usepackage{subfigure}
\usepackage{url}	
\usepackage{xspace}
\usepackage{lscape}

\usepackage{color}

\definecolor{gray}{rgb}{0.7,0.7,0.7}

\usepackage{supertabular}
\usepackage{hhline}
\newcommand\arraybslash{\let\\\@arraycr}
\makeatother
\title{Slime Mould Memristors}
\author{Ella Gale, Andrew Adamatzky and Ben de Lacy Costello}

\date{}                                           

\begin{document}

\maketitle

\begin{abstract}
In laboratory experiments we demonstrate that protoplasmic tubes of acellular slime mould \emph{Physarum polycephalum} show current versus voltage profiles consistent with memristive systems and that the effect is due to the living protoplasm of the mould. This complements previous findings on memristive properties of other living systems (human skin and blood) and contributes to development of self-growing bio-electronic circuits. Distinctive asymmetric $V$-$I$ curves which were occasionally observed when the internal current is on the same order as the driven current, are well-modelled by the concept of active memristors. 
\end{abstract}

\section{Introduction}

Memristors~\cite{Chua_1971} have revolutionised the material basis of computation~\cite{strukov_2008, williams_2012} and  neuromorphic architectures~\cite{Smerieri_2008, Jo_2010,  Pershin_2010, Howard_2012}. Since the announcement of the first documented two-terminal memristor~\cite{strukov_2008} researchers have been eager to experiment with memristors, but they are difficult to synthesize and not yet commercially available. Few looked beyond standard electronic engineering approaches, but those who did uncovered a promising behaviour of living systems. Johnsen et al~\cite{Johnsen_2011} found that conductance properties of sweat ducts in human skin are well approximated by a memristive model~\cite{Johnsen_2011}, and experimental evidence that flowing blood~\cite{Kosta_2011a} and leaves~\cite{Kosta_2011b} exhibits memristive properties was provided by Kosta et al.  In 2008 Pershin et al~\cite{pershin_2008} described an adaptive `learning' behaviour~\cite{saigusa_2008} of slime mould \emph{Physarum polycephalum} in terms of a memristor model.  Memristor theory has been applied to neurons~\cite{84,247} and synapses~\cite{239}, suggesting that memristance might be useful to explain the process of learning.

The plasmodium of \emph{Physarum polycephalum} (Order \emph{Physarales}, class \emph{Myxomecetes}, subclass \emph{Myxogastromycetidae}) is a single cell, visible with an unaided eye. The plasmodium behaves and moves as a giant amoeba. It feeds on bacteria, other microbial creatures, spores and micro-particles~\cite{stephenson_2000}. Structurally \emph{Physarum} is composed of a semi-rigid gel put down by the living protoplasm, a type of cytosol, within it, and this gel is covered with a protective `slime' which gives rise to the plasmodium's colloquial name `Slime mould'. This protoplasm contains many nuclei which can be described as interacting oscillators. Furthermore, the protoplasm undergoes shuttle-transport, switching direction approximately every 50 seconds. Thus, when measuring electrical properties of \emph{Physarum}, we must take care to separate the effects of the living protoplasm, and non-living exteriour gel and slime layer, as well as being aware that the cytosol is a moving and living system which will change over time. 

In an environment with distributed sources of nutrients the plasmodium forms a network of protoplasmic tubes connecting food sources. The network of protoplasmic tubes developed by \emph{Physarum} shows some signs of optimality in terms of shortest path~\cite{nakagaki_2000} and proximity graphs~\cite{adamatzky_ppl_2008}. In \cite{adamatzky_physarummachines} we have used \emph{Physarum} to make prototypes of massively-parallel amorphous computers --- \emph{Physarum} machines ---  capable of solving problems of computational geometry, graph-theory and logic. \emph{Physarum} machines implement morphological computation: given a problem represented by a spatial configuration of attractants and repellents the \emph{Physarum} gives a solution by patterns of its protoplasmic network. This limits application domain of \emph{Physarum} processors to computational tasks with natural parallelism. If we were able to make electronic devices from living \emph{Physarum} we would be able to construct a full spectrum of computing devices with a conventional architecture. Furthermore, to  make a set of `all possible computing devices' it is enough to make a material implication gate and memristors naturally implement material implication~\cite{242,Spc2}. Thus answering the question `Is \emph{Physarum} a memristor?' is a task of upmost priority. 

\emph{Physarum}'s growth can be directed with chemo-attractants and repellants~\cite{delacycostello_2013} and it chooses efficient paths between food sources. Thus, \emph{Physarum} could be used to `design' efficient circuits. Previous preliminary work~\cite{mayne_2013} has shown that \emph{Physarum} can take-up iron-based magnetic particles, so it could be used to lay down efficient circuits and, if the magnetic effects were detrimental to the Physarum, we might expect it to lay down circuits with a good electromagnetic profile, thus, we also investigated the electrical properties of the \emph{Physarum} tubes with and without these particles.

\section{Methods}

\begin{figure}[!tbp]
\centering
\includegraphics[width=0.5\textwidth]{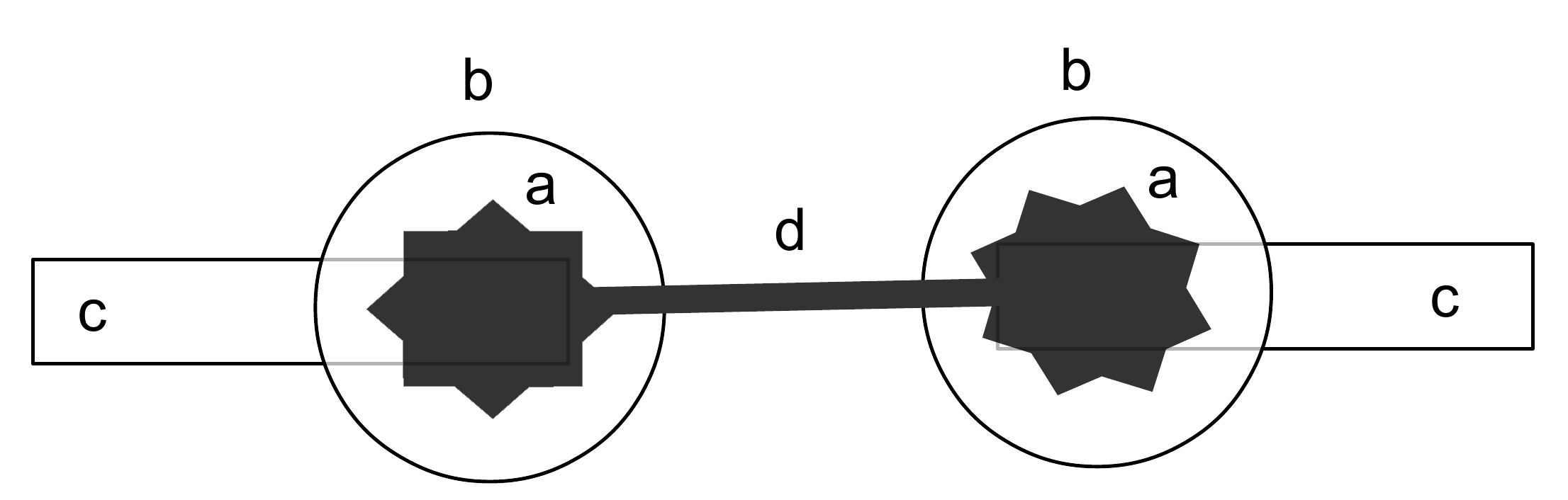}
\caption{A scheme of experimental setup: (a)~Physarum, (b)~agar islands, (c)~electrodes, (d)~protoplasmic tube. All parts of \emph{Physarum} 
shown in dark grey form a single cell.}
\label{scheme}
\end{figure}


Plasmodium of \emph{Physarum polycephalum} were cultivated on wet absorbent paper (to keep the humidity level high) in an aerated, dark environment and fed with oat flakes. The culture was periodically replanted to a fresh substrate. 

The experimental set-up is shown in figures~\ref{scheme}. Two electrodes (Fig.~\ref{scheme}c) were stuck to a plastic Petri dish $\approx$10mm apart and two islands of 2ml agar (Fig.~\ref{scheme}b) were placed on each electrode. To perform the experiments, a \emph{Physarum}-colonised oat-flake was inoculated on one island with a fresh oat flake on the other: \emph{Physarum} would then colonise the other island (Fig.~\ref{scheme}a), linking both electrodes with a single protoplasmic tube (Fig.~\ref{scheme}d). This experimental setup has proved to be efficient in uncovering patterns of electrical activity of \emph{Physarum}~\cite{adamatzky_jones_2011} and \emph{Physarum}'s response to chemical, optical and tactile stimulation~\cite{adamatzky_tactilesensor, adamatzky_RGBsensor, adamatzky_bristles, whiting_2013}.

Electrical measurements were performed with a Keithley 617 programmable electrometer which allows the measurement of currents from pA-3.5mA. Measurements were performed with a voltage range of $\pm$ 50mV (sample 1-13) or 100mV (samples 15-22), a triangular voltage waveform and a measurement rate of 0.5s, 1s or 2s: this is the D.C. equivalent to an A.C. voltage frequency of 2mHz, 1mHz or 0.5mHz. As \emph{Physarum} is a living system and can respond, we compared first and second runs across different dishes. The tests were divided into two batches: batch 1 was measured with timesteps of  0.5s, 1s or 2s and a voltage range of $\pm$50mV and $\pm$100mV; batch 2 was measured only with the 2s timestep and a voltage range of $\pm$100-$\pm$250mV.  

Three different electrode set-ups were tested: thick (2mm) aluminium wire, thin (0.5mm) silver wire and thin aluminium mesh. \emph{Physarum} was also tested for the electrical effect of the uptake of magnetic particles (fluidMAG-D,100nm, 25mg/ml, Chemicell) by inoculating the source or target oat flakes with particles. As \emph{Physarum} is sensitive to light, all tests were run in the dark.


The starting resistance, $R_0$, the hysteresis $H$ (calculated as in~\cite{G1}) and scaled hysteresis, $\bar{H}$, (calculated as a ratio of $R_0$ as in~\cite{Georgiou}) were calculated for the \emph{Physarum} that exhibited hysteresis. To control for experimental set-up variation, $R_0$, $H$ and $\bar{H}$ values were compared to the length of the tube $L$, and the electrode separation. The tube lengths were measured from photographs of the first batch of measurements.

\section{Results}

23 samples were tested; in 1 sample \emph{Physarum} did not form a tube across the electrodes and in another  the \emph{Physarum} grew, formed and abandoned the tube before it could be measured: this sample was used as a control for the gel part of the tube. The remaining 21 samples were measured and these comprised of two samples with magnetic nanoparticles on the inoculation electrode, two samples with magnetic nanoparticles on the target electrode the rest were normal \emph{Physarum}. Three samples were electrically unconnected. 

\subsection{Memristive Effects}

\begin{figure}[!tbp] 
\centering
\subfigure[]{\includegraphics[width=0.49\textwidth]{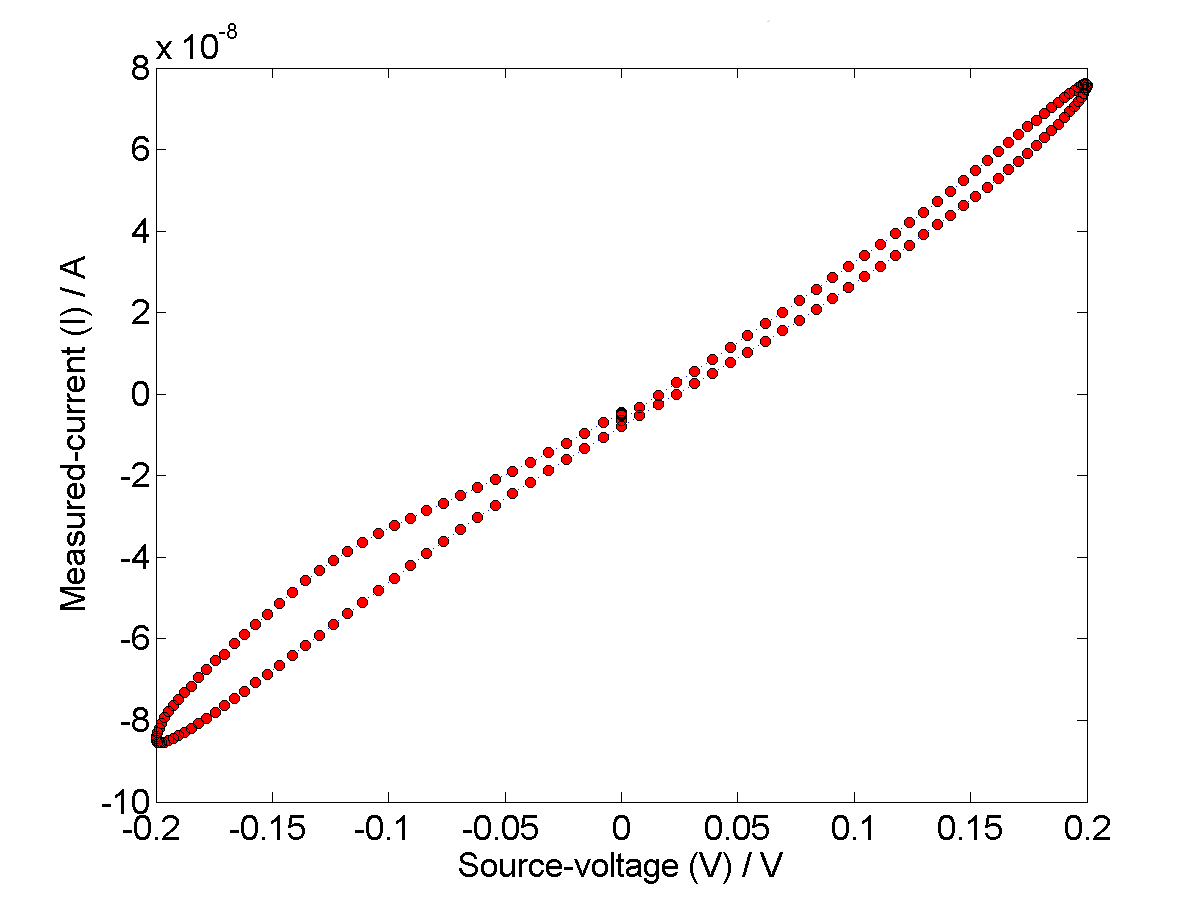}}
\subfigure[]{\includegraphics[width=0.49\textwidth]{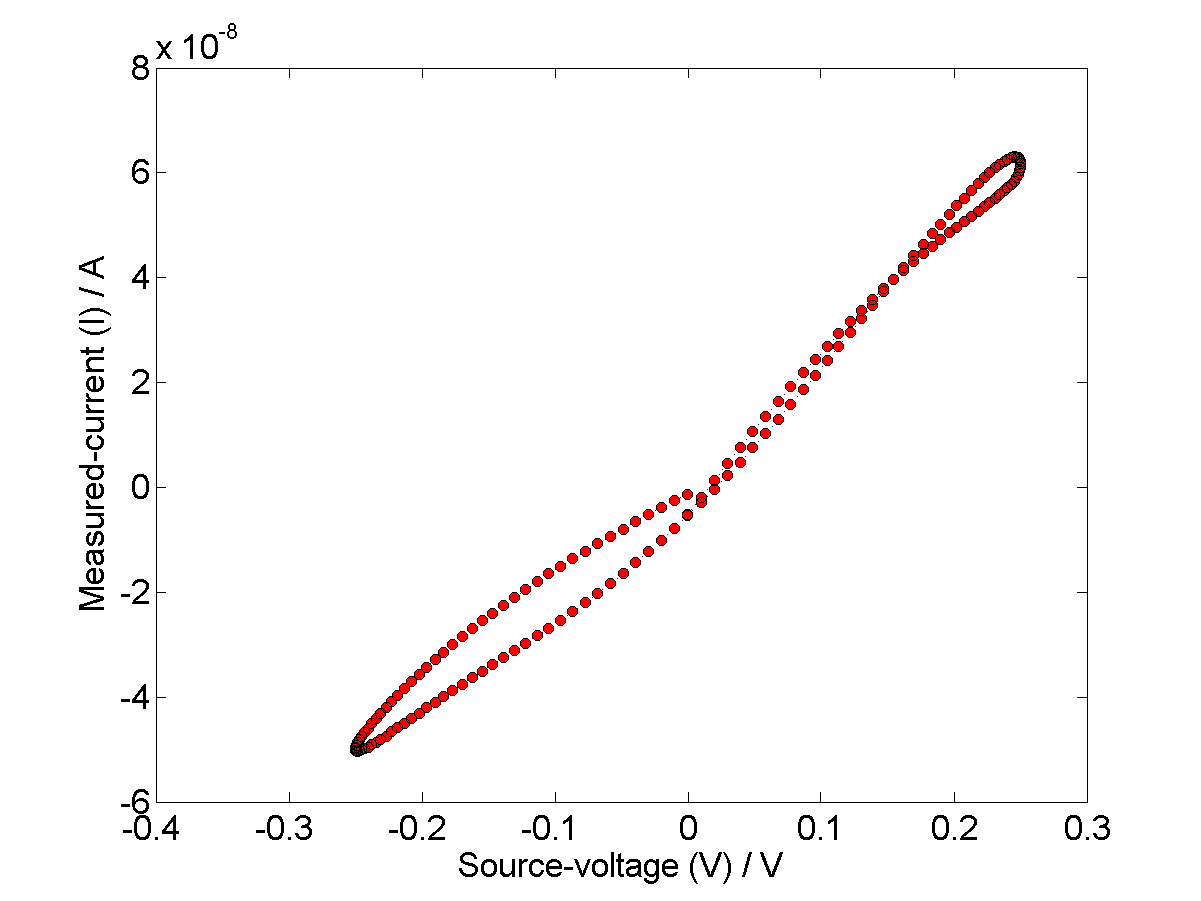}}
\subfigure[]{\includegraphics[width=0.49\textwidth]{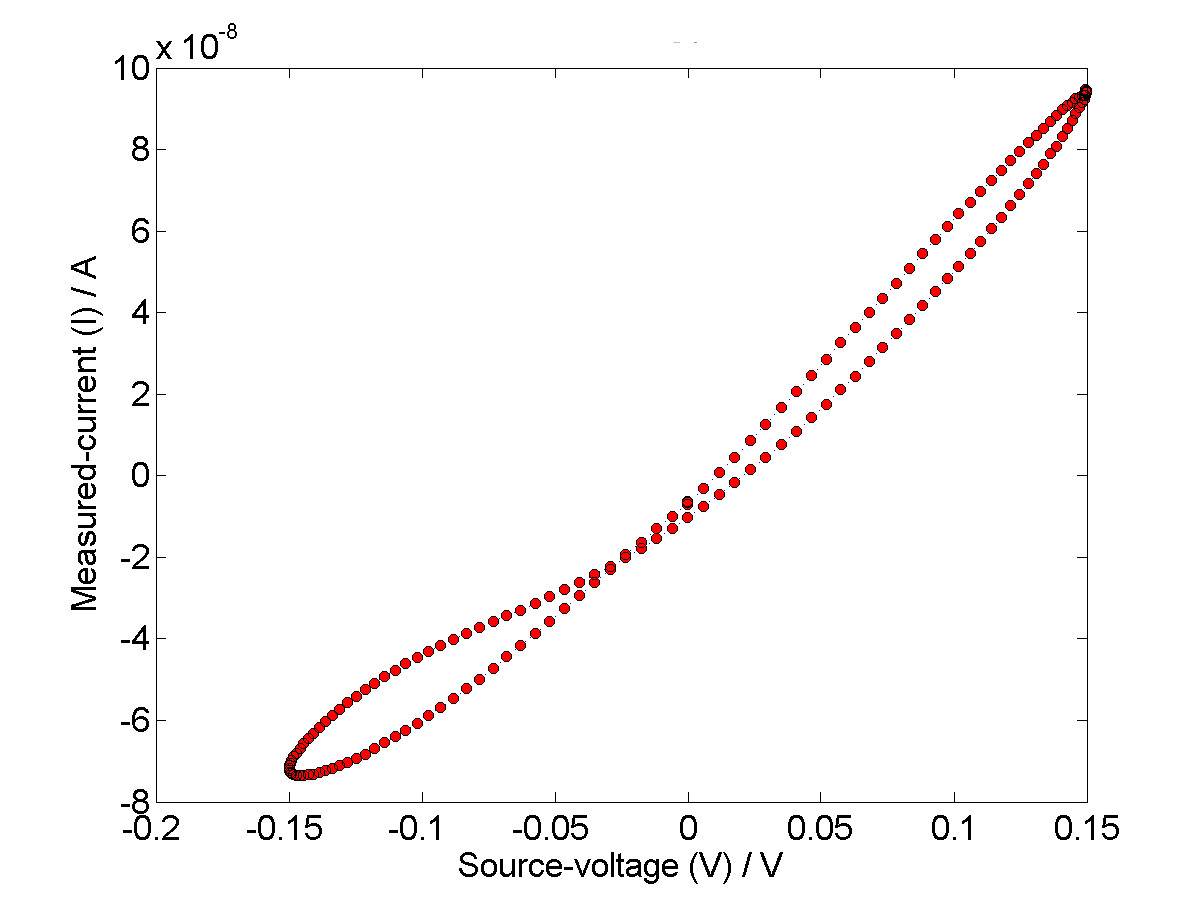}}
\subfigure[]{\includegraphics[width=0.49\textwidth]{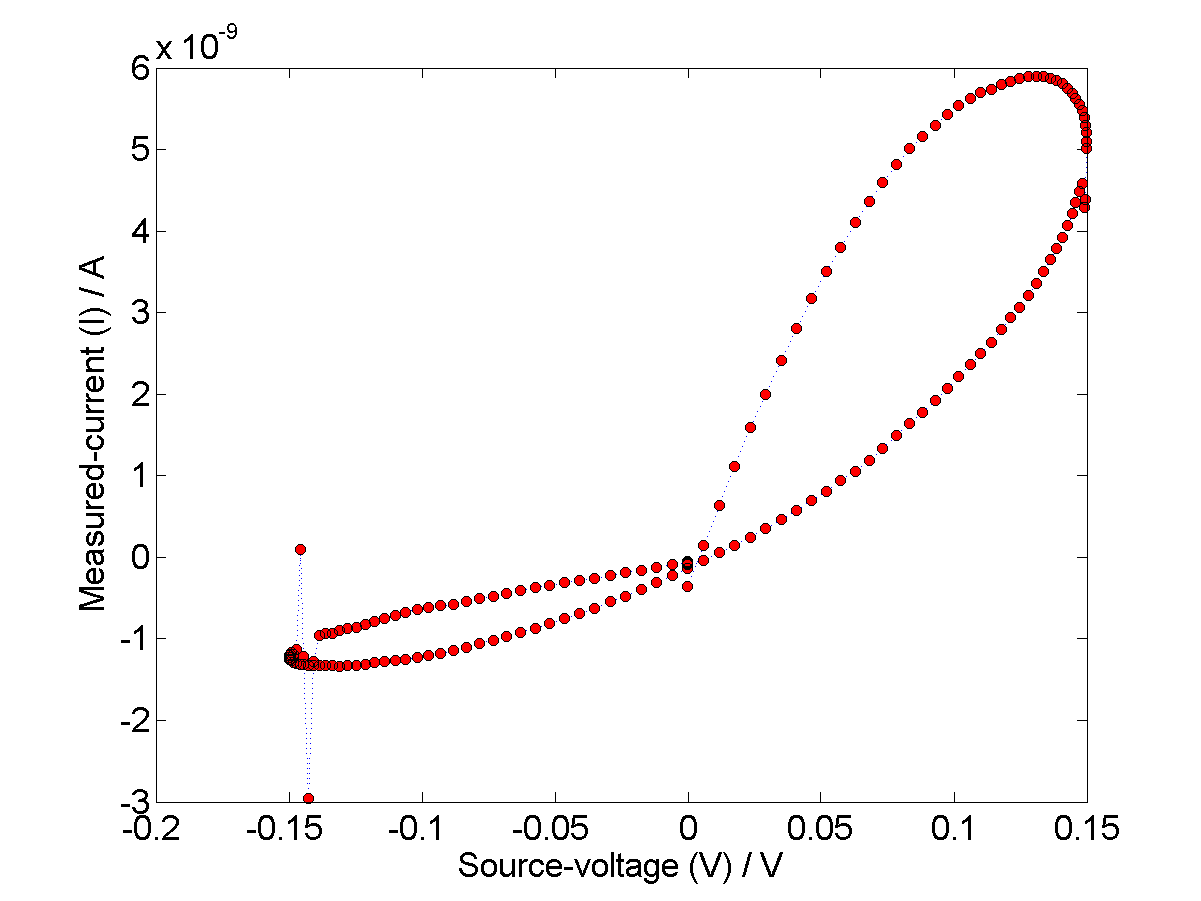}}
\caption{Typical current versus voltage profiles recorded in laboratory experiments with slime mould \emph{P. polycephalum}. }
\label{graphs}
\end{figure}

Of the 11 samples in batch 1, 2 exhibited good memristance curves, the other 9 exhibited open curves as shown in figure~\ref{fig:freq} (these open curves are very similar to those seen in low-voltage TiO$_2$ sol-gel memristor measurements which are indicative of memristive results at high voltage measurements~\cite{Gale_ElectrodeMetal}). For batch 2, which were measured at a larger voltage range (between $\pm100$mV and $\pm$250), 8 out of 8 samples showed good memristive curves as shown in figure~\ref{graphs}. 

Comparison between the memristance curves shown in figure~\ref{graphs} and the results for an abandoned tube shown in figure~\ref{fig:abandoned} shows that the memristance is due to the living \emph{Physarum} protoplasm. Similar to inorganic memristors, this could be due to voltage-driven charge transport. Shuttle transport reverses direction around every 50s and this could give rise to a measurable hysteresis, however because these data were measured at $\Delta t \approx$2s for 160 steps the period is 320s which is over 3 times the period of the shuttle transport and, as such, is not the cause of the measured effect. Longer time-scale current responses have been observed in d.c. experiments~\cite{Gale_Light} which could be related to the memristive effect. Finally, it could be due to a voltage-mediated change in material properties of the protoplasm which has a relaxation time, leading to a memristive hysteresis -- this seems to be the mostly likely explanation because repeated applications of voltage 
increased the resistance (see figure~\ref{fig:freq}).

No electrical effect of the different electrode types was seen beyond the mechanical difficulties: the thicker aluminium electrodes put strain on the tube when measured leading to breakages or short-circuits. No discernible electrical effect was seen from the presence of magnetic nanoparticles. The \emph{Physarum} picked them up and internalised them, but the measured resistance curves were qualitatively the same as those without the particles and within the same range. Our other work~\cite{mayne_2013} shows that the nanoparticles are localised within the gel part of the \emph{Physarum}, so this lack of effect is not because the nanoparticles are not present. One possibility could be that the \emph{Physarum} might internalise or biofoul the nanoparticles. 

This effect whereby open loops at low voltages show memristance at high voltages suggests that, like TiO$_2$ sol-gel memristors, the low-voltage open-loop behaviour is related to memristance. These results show that relatively large voltages are needed for the measurement of memristance in \emph{Physarum}.

\subsection{Repeatability and the effect of frequency}

The testing process may have resulted in behavioural modification of the \emph{Physarum}: successive applications of voltage caused the \emph{Physarum} to abandon the tube as observed by a thinning and lightening of the tube. However \emph{Physarum} was still alive and active after testing, and the tube abandonment could be due to it exploring the environment for other food sources (the time spent connecting the two oat-flakes was commensurate with that observed without electrical testing, but due to the high variance in behaviour we cannot say if the applied voltage harmed the \emph{Physarum}). 

\begin{figure}[!tbp]
 \centering
 \includegraphics[width=0.6\textwidth]{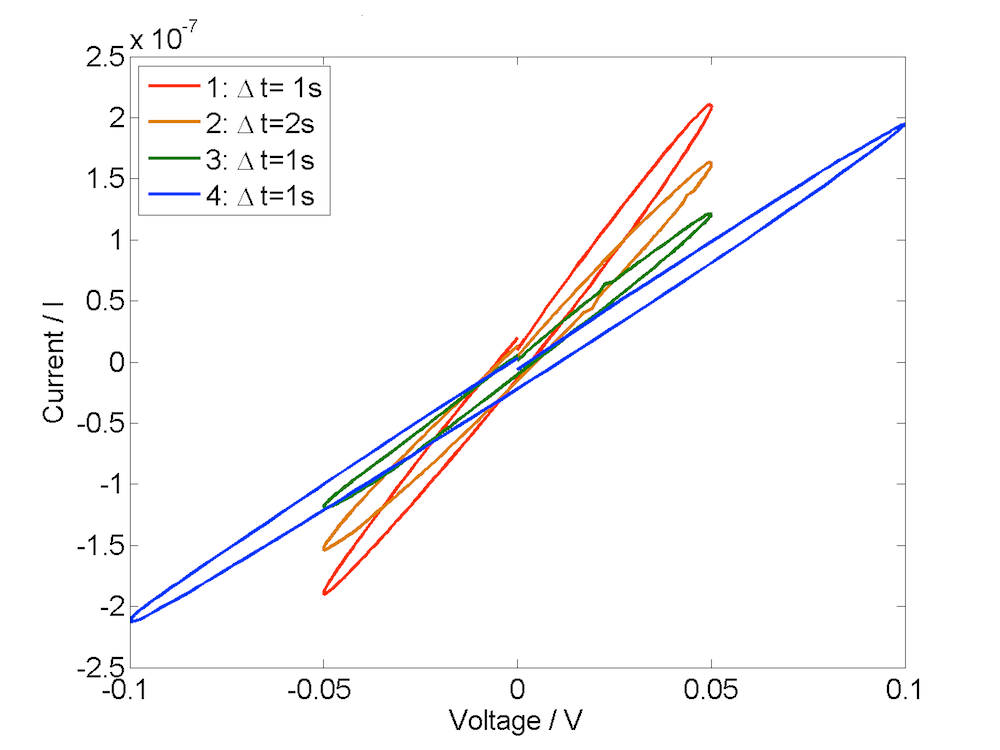}
 \caption{Open-curves measured under different frequencies}
 \label{fig:freq}
\end{figure}



\begin{figure}[!tbp]
 \centering
 \includegraphics[width=0.5\textwidth]{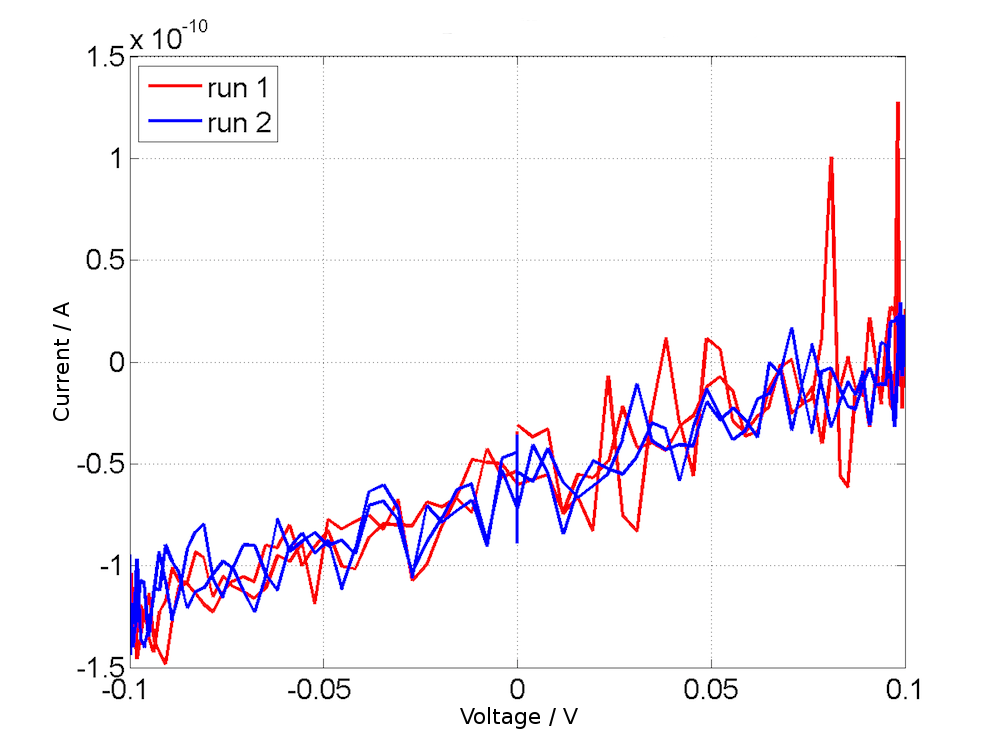}
 \caption{Repeated I-V runs performed on an abandoned tube. No memristance or increasing resistance effect is seen.}
 \label{fig:abandoned}
\end{figure}

As frequency can effect the size of memristor hysteresis (due to natural response speed of the system) the voltage waveform frequency was altered. As increasing the voltage range can turn the `open-loop' type of memristor into a pinched hysteresis loop, the lack of pinched hysteresis on the open-loop memristor responses could be due to the chosen voltage range. The effect of frequency is tested in Fig.~\ref{fig:freq}, where three repeats of the same $\pm$V range is tried at the standard and twice the frequency and one larger $\pm$V range is tried at the standard frequency. The shape is qualitatively similar over this voltage range and unaffected by frequency over this voltage range. Figure~\ref{fig:freq} shows that repeated applications of voltage causes the the resistance of the tube to increase (similar results were observed on the two repeats with other samples). 

As Fig.~\ref{fig:abandoned} shows no repeated resistance change for an empty tube, this suggests that the protoplasm part of \emph{Physarum} is the material responsible for the observed memristance rather than a chemical or physical change in the structure of the outer parts of the tube.

\subsection{Study of memristor properties}

\subsubsection{The Effect of Electrode Separation and Tube Length}

The range of tube lengths found was 6.25mm to 43mm, with a mean of 19.71mm and standard deviation of 10.64mm: this high deviation is because the \emph{Physarum} initially explores the space following a chemical gradient to connect the oat flakes. The tube length shortens over time as the \emph{Physarum} increases the efficiency of connection between food sources, for example, the length of one protoplasmic tube went from 5mm to 4.26mm over a day. No correlation was found between: $R_0$ and $L$; $\bar{H}$ and $L$; $H$ and $L$; $H$ and $R_0$; $R_0$ and Electrode separation (graphs not shown), this demonstrates that the tube length ($L$) or electrode separation cannot be used to control the electrical properties  and that the starting resistance ($R_0$) is not a predictor for the hysteresis ($H$ and $\bar{H}$): as a comparison, results for sol-gel memristors are given in~\cite{M1}. The total power used over an $I$-$V$ loop and the average power used were also calculated for 11 samples, no correlation was found between the power and $R_0$, or the power and electrode separation (graphs not shown). Thus we can conclude that the variation in electrical response is due to the variation between individual samples of \emph{Physarum} and not the variation in set up or tube length.

\section{Theoretical Analysis of the Slime Mould as an Active Memristor}


\begin{figure}[!tbp]
\centering
 \subfigure[]{\includegraphics[width=0.49\textwidth]{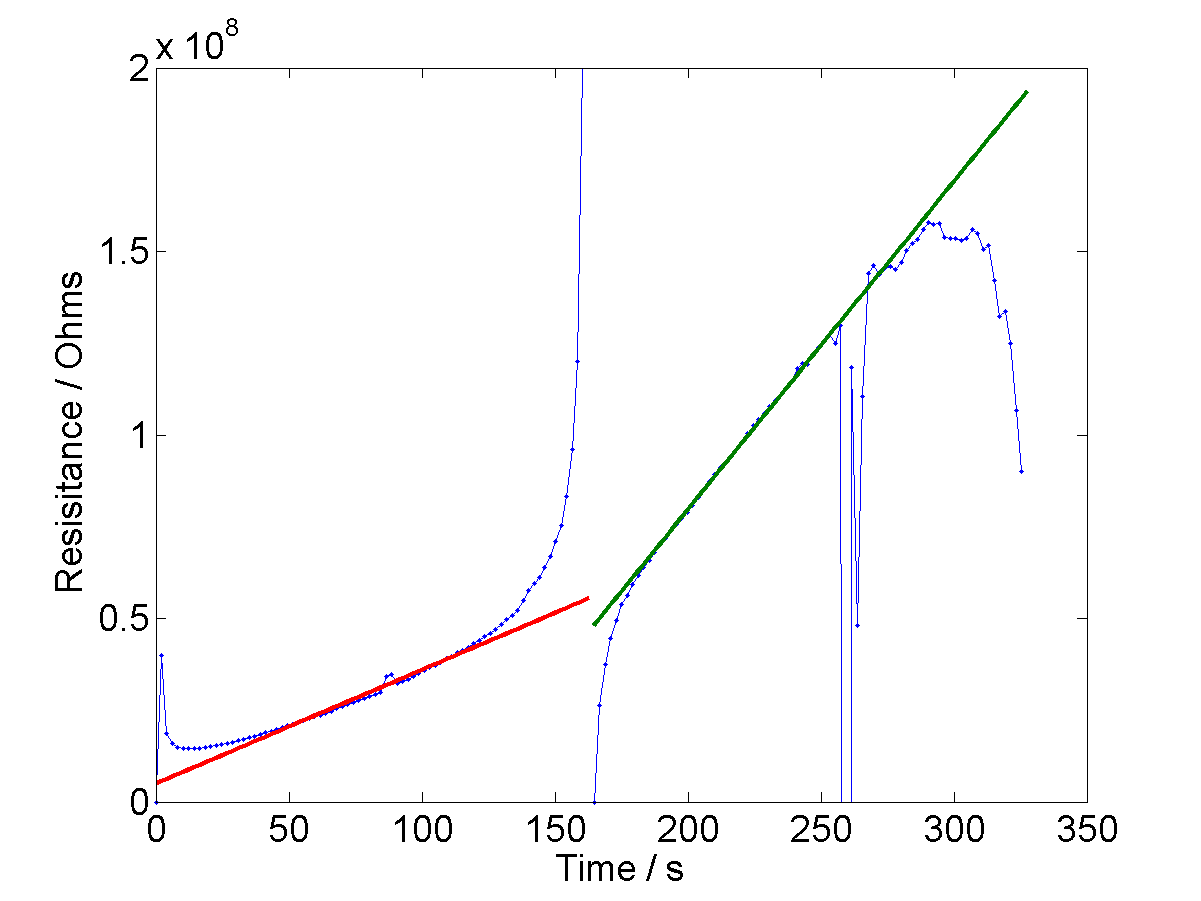}}
 \subfigure[]{\includegraphics[width=0.49\textwidth]{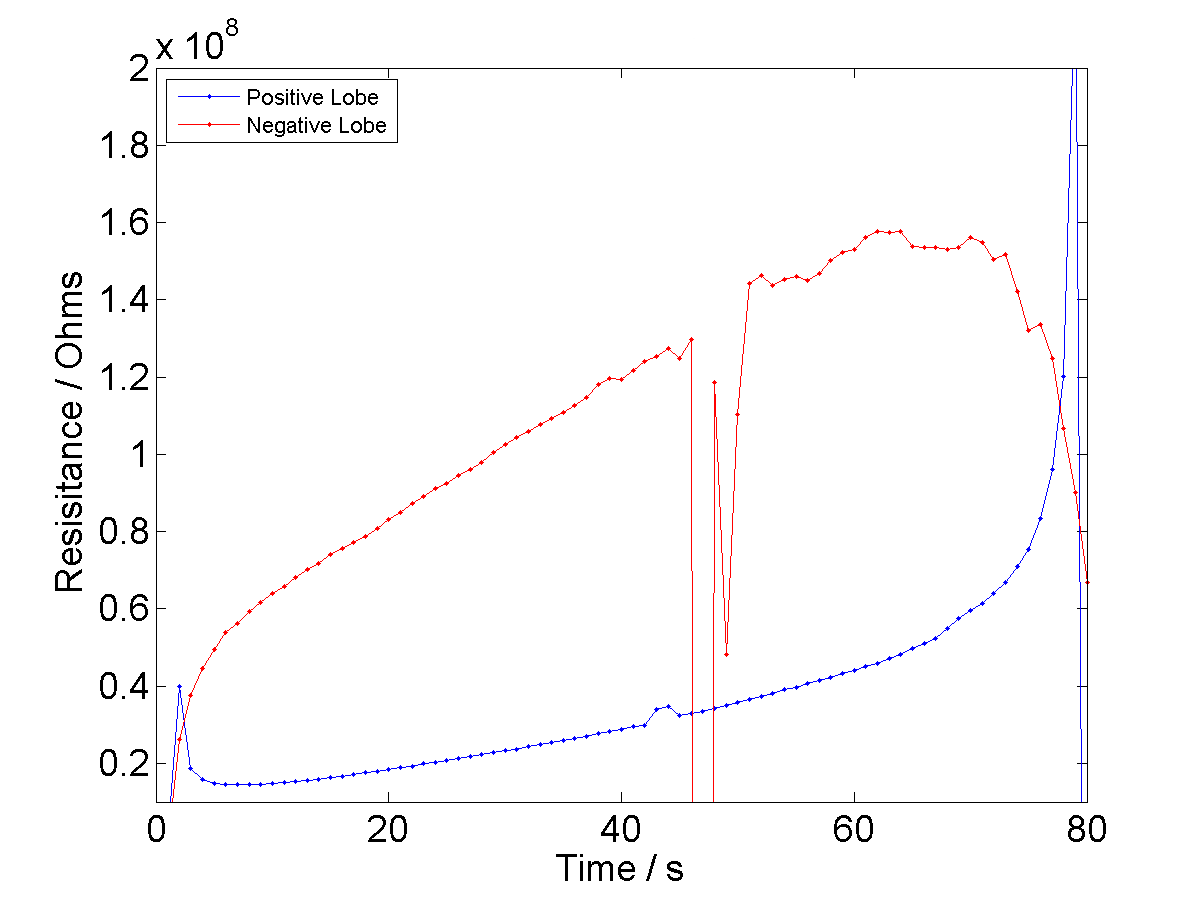}}
\caption{Instantaneous Resistance - time plots. 
(a)~With tangents,
(b)~As a closed loop in resistance-time space}
\label{fig:R-t}
\end{figure}

The shape of the curve in figure~\ref{graphs}d is interesting, as it shows asymmetry between in the resistance change rate, and has not been observed in our inorganic memristors. The memory-conservation theory of memristance~\cite{mem-con} explains memristive effects in terms of an interaction between state-carrying ions, $q$, and conduction (state-sampling) electrons $e^{-}$. To undergo locomotion, \emph{Physarum} exhibits shuttle transport where the protoplasm is moved backwards and forwards: ions in the protoplasm would be moved around by this motion and this gives could give rise to a background current. Thus, this background current should be included and we investigated this in order to try and understand the shape of the distinctive \emph{Physarum} memristor (see figure~\ref{graphs}d). A similar approach has been used to model ReRAM, where the electromotive force associated with a `nanobattery' is added to a memristor circuit~\cite{Waser}.

\begin{figure}[!tbp]
\centering
 \includegraphics[width=0.49\textwidth]{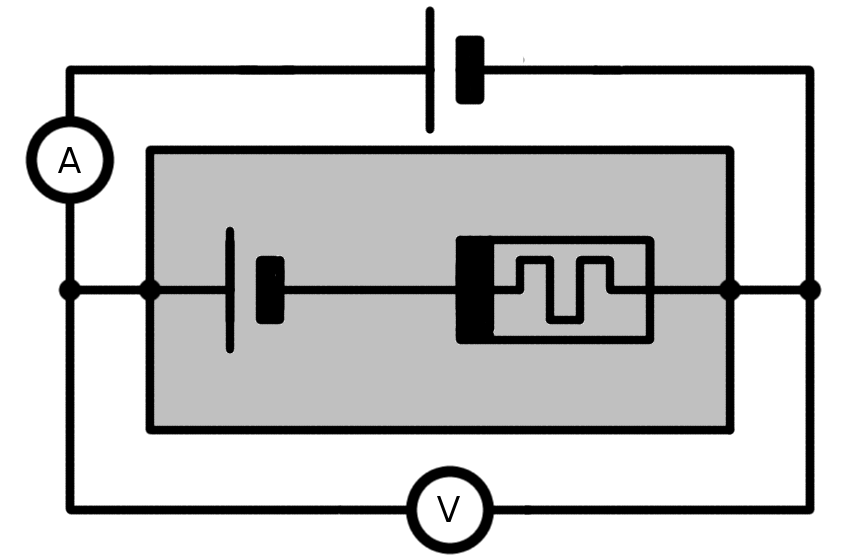}
\caption{The equivalent circuit for the \emph{Physarum} memristor, the shaded area represents the middle active memristor 2-port.}
\label{fig:Circuit}
\end{figure}

Figure~\ref{fig:Circuit} shows a circuit which could be used to model the situation: this circuit contains a 2-port `black box' which we measure. We assume that this 2-port contains a current source (battery) because \emph{Physarum} is alive and uses chemical energy to produce reactions and the motion of the membrane, and a memristor (or a memristor-resistor in series). In fact, from long-term experiments we have seen a slow oscillation with a half-period of around 700~\cite{Gale_Light} that could fit the description of such a current source, especially as it was observed at $\approx 2 \times 10^{-9}$A, so it is on the same order as our lower current $V$-$I$ curve measurements. With the addition of an internal current source, we are now modelling the \emph{Physarum} as an active memristor~\cite{active} (standard memristors are passive components). 

The current, $i_q$, is the background current from the living \emph{Physarum}. From the circuit in figure~\ref{fig:Circuit} we can write the following expression for the measured current, $I_{tot}$:

\begin{equation}
I_{tot} = i_d \pm i_q \, 
\label{eq:Itot}
\end{equation}

where $i_d$ is the current that is driven by the external voltage, $V$. The background battery can either add to or oppose the external power source, and thus the background current is either in the same direction or opposite direction to the driven current, as we use $+$ and $-$ to represent this, where it is understood that the internal ions may not have the same charge as the electrons and we take $+i_q$ to be the direction of increase in total current. $V_q$ is the voltage associated with the internal `battery'. 

If we have a current that adds to our driven current at one point in time and subtracts at another, we would expect to see a non-rotational symmetric memristor curve. 

From equation~\ref{eq:Itot} we can write the memristive response, $R$, as:
\begin{equation}
R = M \pm \frac{V_q}{i_q} \;,
\label{eq:R}
\end{equation}
 
where $M$ is the memristance due to the motion of ions under the applied voltage --- that expected from the memory-conservation theory of memristance --- and the second term is the internal resistance response due to the background current, which we label as $M_{int}$. 

We don't know what the form of $i_q$ is, but there are two options, we can model it as a sine wave with a period of roughly 700s, or we can model it as a bipolar piece-wise linear (BPWL) waveform, which corresponds more to what is observed down a microscope when watching \emph{Physarum} shuttle transport. As the $I$-$V$ curves took a total of $\approx$350s to run, we can model the current as being constant over this period, especially if it is the long-time oscillations observed in~\cite{Gale_Light} and thus additive to the memristance current in one direction and subtractive in the other. 

We can descretise equation~\ref{eq:R}, to get an expression for the descretised rate of change of memristance, $\frac{\Delta M}{\Delta t}$ as:
\begin{equation}
\frac{\Delta R}{\Delta t} (+) = \frac{\Delta M}{\Delta t} + \frac{\Delta M_{int}}{\Delta t} \; ,
\end{equation}
for the positive lobe of the plot and 
\begin{equation}
\frac{\Delta R}{\Delta t} (-) = \frac{\Delta M}{\Delta t} - \frac{\Delta M_{int}}{\Delta t} \; ,
\end{equation}
for the negative lobe of the plot, see figure~\ref{fig:R-t}.

Assuming that the rate of change of the $M(t)$ does not change over the memristors range although it does change direction, (which is an approximation), we can substitute for $\frac{\Delta M}{\Delta T}$ and write
\begin{equation}
g \frac{\Delta R}{\Delta t}(+) + \frac{\Delta M_{int}}{\Delta t} = \frac{\Delta R}{\Delta T}(+) - \frac{\Delta M_{int}}{\Delta t} \:,
\label{eq:gR+}
\end{equation}
where $g$ is the factor that $\frac{\Delta R}{\Delta t} (-)$ is bigger than $\frac{\Delta R}{\Delta T} (+)$ by and it is equal to 2.88, i.e. the rate of change of resistance is around 3 times faster on the negative lobe compared to the positive, leading to a non-rotationally-symmetric pinched hysteresis loop.

We can calculate the actual rates $\frac{\Delta R}{\Delta T} (+)$ and $\frac{\Delta R}{\Delta T} (-)$ from the measured current $I$, which we do by calculating the `instantaneous' memristance $M(\Delta t)$ at each measurement point:
\begin{equation}
M(\Delta t) = \frac{\Delta V_s}{\Delta I} \;, 
\end{equation}
and this is shown in figure~\ref{fig:R-t}. Around zero and small values of $V$ we get large discrepancies, due to the method we're using to calculate the memristance, but over most of the curve we can see that the straight-line approximation of the change in memristance holds pretty well. Figure~\ref{fig:R-t}b shows that there are two gradients, a shallower one for the positive loop and a  steeper one for the negative loop. If these gradients were equal we would have a standard (ideal) memristor curve. 

The memristor curve is commonly broken up into 4 segments: 1: $0V \rightarrow + \mathrm{max}(V_s)$; 2: $+ \mathrm{max}(V_s) \rightarrow 0V$; 3:$0V \rightarrow  - \mathrm{max}(V_s)$; 4:$- \mathrm{max}(V_s) \rightarrow 0V$. We chose to fit a straight line to the 1$^{\mathrm{st}}$ and 3$^{\mathrm{rd}}$ segments as they start from the same place (0V), these lines are shown on the curve and their equations are gradients of 3.1009$\times10^{5} \Omega s^{-1}$ and 8.9348$\times10^{5} \Omega s^{-1}$, $y$-intercepts of 4.9696$\times10^{5} \Omega$ and -9.9034$\times10^{7} \Omega$ (the negative intercept is obviously unphysical and is a result of approximating and changing $\frac{\Delta R}{\Delta t}$ by a tangent) with a norm of residuals of 5.7685$\times10^{6} \Omega$ and 5.8028$\times10^{6} \Omega$ segments 1 and 3 respectively. 
                      
We can get a measure of the memristance of the cell's `internal battery' from rearranging equation~\ref{eq:gR+}:\begin{equation}
\frac{\Delta M_{int}}{\Delta t} = \frac{\left( g-1 \right) \frac{\Delta R}{\Delta t}(+)}{-2} \; ,
\end{equation}
                
this gives us a negative slope of -2.91485$\times 10^{5}\Omega s^{-1}$ with a negative resistance intercept whose modulus is 94\% $R_0$ (where we are taking $R_0$ as the $y$-intercept from the fitted tangent for the first segment). Negative resistance implies the presence of active components in our test circuit, verifying our approach of treating the cells as possessing an `internal battery'. This shows that, at these voltages (which are close to physiological voltages), the cell's internal `battery' gives physiological currents close to our driven current. Thus, to model living cells over physiological ranges, it seems that active memristors are a better approximation than passive memristors.

%


\section{Discussion}

The results clearly show hysteresis and memristive effects in \emph{Physarum Polycephalum}. The frequency and voltage range choice effected the results, we found that a timestep of $\delta t=2s$ and a $\pm$ of over 200mV gave the best results. At low voltages, an open-curve shape was measured instead, which we suspect is the memristance effect when measured at below a threashold voltage. As the memristive effect disappeared when the \emph{Physarum} moved elsewhere, and an abandoned tube showed a high linear resistance, we conclude that the memristive response is due to the living protoplasm. Active memristor models show a promising explanation for the asymmetric shape seen when the memristor current response is below $10^{-8}$A, for higher current responses the internal current is small enough that ignoring it and modelling the \emph{Physarum} as an ideal memristor is a valid approximation.

It is intriguing that \emph{Physarum} exhibits memristive ability, given that it is a simple biological system that is nonetheless capable of habituation and learning and that neurological components (synapses, ion pumps) also exhibit memristance and learning abilities. This could suggest that evolution may have made use of memristance in learning systems. The presence of biological active memristors suggests that biological chaotic circuits could be possible (active memristors are a common component of chaotic circuits~\cite{active}), and even that they may have been utilised by evolution.

Current versus voltage profiles measured for protoplasmic tubes of \emph{P. polycephalum} exhibit great variability in the magnitude of hysteresis and the location of pitch points. This is to be expected given the fact that slime mould is an ever-changing living entity and although attempts were made to standardise the experimental set up such as the measurement of single protoplasmic tubes across a known electrode gap it proved difficult to precisely control the morphology of the tubes. For example even though the electrode distance can be controlled this does not ensure standardisation of the protoplasmic tubes length or the width. It also proves difficult to control the position and numbers of small sub-branches which may arise during experimental measurements. Although these do not usually contact the electrode except for sub branching at the terminal ends, this alteration in morphology is likely to affect the conductivity. Thus future research would focus on stabilisation of protoplasmic tubes. 

Stabilisation could be achieved by employing the \emph{Physarum}'s potential for internalisation and re-distribution of conductive and magnetic particles~\cite{mayne_2013} or at least constraining the growing with some kind of scaffolding~ \cite{Tian_2012}. There is also the potential that the slime mould could construct an internal scaffold or that this could be induced by appropriately applied external fields, in fact morphological control could be accomplished by application of appropriate fields per se~\cite{delacycostello_2013}.  Despite the complications of living electronics, we have demonstrated that it is feasible to implement living memristive devices from slime moulds \emph{P. polycephalum}. 

We believe that future electronic designs will incorporate growing slime mould networks capable of forming a skeleton of conductive  information processing elements as part of integrated computing circuits. The slime mould circuits will allow for a high density of computing elements and very low power consumption. To date the useful lifetime of a slime mould memristor is 3-5 days. However, future studies on loading and coating of the tubes with functional materials with a dual role of structural re-enforcement such as nano-metallic, nano-magnetic or nano-structured semiconducting particles, conducting polymers etc. should enable us to increase the life span substantially whilst also imparting a diverse range of tunable electronic characteristics. If electronic circuits can be `grown' or laid-down from \emph{Physarum}, it would be very useful for reducing the water requirements and poisonous waste of the electronics industry.



\end{document}